\documentclass[pra,twocolumn,showpacs,letterpaper,showpacs,superscriptaddress]{revtex4}
\usepackage{amsfonts}
\usepackage{amsmath}
\usepackage{amssymb}
\usepackage{graphicx}
\usepackage[usenames,dvipsnames]{color}
\newcommand{\la}{\langle}
\newcommand{\ra}{\rangle}
\newcommand{\be}{\begin{equation}}
\newcommand{\ee}{\end{equation}}
\newcommand{\bea}{\begin{eqnarray}}
\newcommand{\eea}{\end{eqnarray}}
\newcommand{\om}{\omega}
\newcommand{\Om}{\Omega}

\newcommand{\pa}{\partial}
\newcommand{\br}{{\bf r}}
\newcommand{\bx}{\hat{{\bf x}}}

\newcommand{\bp}{\hat{{\bf p}}}

\newcommand{\bK}{\hat{\bf K}}
\newcommand{\hK}{{\hat{K}}}

\newcommand{\eps}{\epsilon}

\newcommand{\bE}{\hat{\bf E}}
\newcommand{\bF}{\hat{\bf F}}
\newcommand{\bH}{\hat{\bf H}}

\newcommand{\bD}{\hat{\bf D}}
\newcommand{\bB}{\hat{\bf B}}

\newcommand{\NN}{{\cal N}}
\newcommand{\hx}{\hat{x}}
\newcommand{\hp}{\hat{p}}
\newcommand{\hF}{\hat{F}}
\newcommand{\hE}{\hat{E}}
\newcommand{\cE}{\mathcal{E}}

\preprint{}

\begin{document}

\title{Electromagnetic Energy, Absorption, and Casimir Forces. Inhomogeneous Dielectric Media}

\author{F. S. S. Rosa}
\affiliation{Theoretical Division, MS B213, Los Alamos National Laboratory, Los Alamos, NM 87545, USA}
\affiliation{Laboratoire Charles Fabry, Institut d'Optique, CNRS, Universite Paris-Sud, Campus Polytechnique, RD128, 91127, Palaiseau Cedex, France}

\author{D. A. R. Dalvit}
\affiliation{Theoretical Division, MS B213, Los Alamos National Laboratory, Los Alamos, NM 87545, USA}

\author{P. W. Milonni}
\affiliation{Theoretical Division, MS B213, Los Alamos National Laboratory, Los Alamos, NM 87545, USA}
\affiliation{Department of Physics and Astronomy, University of Rochester, Rochester, NY 14627, USA} 

\begin{abstract}
A general, exact formula is derived for the expectation value of the electromagnetic energy density of an inhomogeneous absorbing and dispersive dielectric medium in thermal equilibrium, assuming that the medium is well approximated as a continuum. From this formula we obtain the formal expression for the Casimir force density. Unlike most previous approaches to Casimir effects in which absorption is either ignored or admitted implicitly through the required analytic properties of the permittivity, we include dissipation explicitly via the coupling of each dipole oscillator of the medium to a reservoir of harmonic oscillators. We obtain the energy density and the Casimir force density as a consequence of the van der Waals interactions of the oscillators and also from Poynting's theorem.
\end{abstract} 


\maketitle

\section{Introduction}
Analyses of quantum fluctuations of the electromagnetic field in dielectric media typically take a macroscopic approach in which the field is quantized under the assumption that the medium is characterized by an electric 
permittivity \cite{fieldquant}. Much of this work is restricted to fields at frequencies for which absorption is negligible, and some of it assumes furthermore that dispersion is negligible. An important exception is the work of Huttner and Barnett \cite{huttnerbarnett}, who allow for both absorption and dispersion; as in virtually all work in this area, they treat the (unexcited) atoms of a homogeneous dielectric medium \cite{term} as harmonic oscillators \cite{fano}, and dissipation is accounted for by coupling each of these oscillators to a ``bath" of reservoir oscillators. The atoms are also coupled to the electromagnetic field, and the entire system of atoms, reservoirs, and field oscillators is then diagonalized along the lines of Fano's method \cite{fano} to yield expressions for quantized electric and magnetic fields. 

In a less elegant approach one writes Heisenberg equations of motion for the atom, reservoir, and field
variables. The effect on each atom of its coupling to its reservoir is to introduce a damping force and
a Langevin force, the latter ensuring the preservation of canonical commutation relations. The Langevin force on each
atom results in a fluctuating ``noise polarization" of the type introduced in Rytov's theory of fluctuational electrodynamics \cite{Rytov}, and used by 
Lifshitz \cite{lifshitz} in his theory of
the van der Waals-Casimir force per unit area between two dielectric half-spaces. The operator
Maxwell equations with this noise polarization yield expressions for the quantized transverse fields having the same form as  those obtained by Huttner and Barnett \cite{pap1,erber}. 

One purpose of the present paper is to extend earlier work \cite{pap1,pap1typos}, hereafter referred to as I, to the 
case of \emph{inhomogeneous} dielectric media. This leads to a formula for the total energy density of a dispersive, dissipative, and inhomogeneous dielectric medium in terms of the dyadic Green function $G$. We derive the force density at finite temperature $T$ associated with spatial variations of the complex electric permittivity
$\eps(\br,\om)$ of the medium:
\bea
{\bf f}(\br)&=&-\frac{\hbar}{8\pi^2 c^2}{\rm Im}\int_{0}^{\infty}d\om\om^2\coth\Big(\frac{\hbar\om}{2k_BT}\Big)\nabla\eps(\br,\om)\nonumber\\
&&\mbox{}\times G_{ii}(\br,\br,\om).
\label{intro1}
\eea
%
%
$G_{ii}(\br,\br,\om)$ is the sum of the three diagonal components of the Fourier transform of the Green
dyadic in the summation convention for repeated indices used throughout this paper. From (\ref{intro1}) one obtains, for instance, the Lifshitz formula \cite{lifshitz} for the force per unit area between two semi-infinite dielectric media separated by a distance $d$, as was shown in the early work of Dzyaloshinskii \emph{et al.} \cite{igor} and Schwinger \emph{et al.} \cite{julie}, and from that formula one obtains in various limits, as is well known, the Casimir force between perfectly conducting plates, the van der Waals interaction between two atoms, the Casimir-Polder interaction between an atom and a conducting plate, etc. 

The main motivation for the present work is not just to rederive the general force density (\ref{intro1}) by different methods, but rather to obtain it by allowing from the outset for absorption. Derivations of the ($T=0$)
Lifshitz formula based on changes in electromagnetic energy \cite{vankamp}, following Casimir's original 
calculation \cite{casimir}, are based on the zero-point electromagnetic energy $\sum_j\hbar\om_j/2$, the sum being over all possible modes of the field. The frequencies $\om_j$ in such calculations are first determined for the case of nondissipative media (real permittivities), and dissipation is later accounted for by making in effect an analytic continuation based on the requirement from causality that the permittivity is analytic in the upper half of the complex frequency plane. The question of how to apply this approach by admitting dissipation from the start is an old one in the theory of Casimir forces. Agarwal \cite{agarwal}, for instance, remarks that ``if the damping of the dielectric function is included, then the [method based on zero-point energy] seems to fail. In presence of damping the normal-mode frequencies are complex, and it is not clear what one should sum over to obtain the interaction energy." In Reference \cite{pwm}, similarly, it is noted that ``it is not obvious how to extend [the zero-point-energy approach] to the case of absorbing media, where the [permittivities] are complex." 
Ginzburg \cite{ginzburg} observes that, ``Apart from a general maxim that `victors should not be judged' one can justify the derivation of [the Lifshitz formula] on the basis of [the assumption that a mode of frequency $\om$ has zero-point energy $\hbar\om/2$]  for absorbing media as follows. Firstly, the permittivities [appearing in the Lifshitz formula] are functions. Secondly, the function $\eps(\om)$ is always real on the imaginary axis. The result obtained for transparent media ... must therefore clearly be the same as the appreciably more general one which is applicable to absorbing media. However, this would hardly confirm such a conclusion unless it had been obtained earlier without additional assumptions. Both for this reason and also bearing in mind other related or similar problems one must somehow consistently generalize the expansion in eigenoscillations with frequencies $\om_{\alpha}$ ... to absorbing media." Ginzburg proceeds to introduce the concept of (orthogonal) ``auxiliary" field modes associated with the immersion of the entire system ``in some auxiliary resonator with perfectly conducting walls ... The frequency $\om$ is considered to
be a parameter, while the eigenfrequencies of the resonator $\om_{\alpha}$ are determined from the homogeneous field
equations" corresponding to Maxwell's equations without a fluctuating polarization density. 

There are arguably more direct ways of explaining the success of calculations that ``consider directly
only transparent media" and attribute Casimir forces to changes in zero-point energy due to the
presence of material media. A clue in this direction is provided by the case of a homogeneous dielectric medium: the  zero-point energy has the same form regardless of whether one allows ``directly" for absorption \cite{pap1}. 

In the following section we derive the Casimir free energy and force density for a dispersive and absorbing, linear, inhomogeneous dielectric medium based on an extension of the classical formula 
\be
W_c=-\frac{1}{2}\alpha(\om){\bf E}^2(\br)
\label{intro2}
\ee
for the change in energy when an electric field ${\bf E}$ of frequency $\om$ induces an electric
dipole moment $\alpha(\om){\bf E}$ in a particle having a real polarizability $\alpha(\om)$. For an atom in a state characterized by the polarizability $\alpha(\om)$ this is just the quadratic Stark shift of its energy level. The generalization needed to obtain the Casimir free energy only requires allowance for a complex polarizability and the  effect on each dipole of the fields from all the other dipoles. We explicitly include dissipation and Langevin forces resulting from the coupling of the dipole oscillators to their reservoirs and show that the force density obtained in this way is equivalent to that obtained in the seminal work of Dzyaloshinskii \emph{et al.} \cite{igor} using diagrammatic methods and subsequently by other authors by various other techniques. 
In Section \ref{sec:macro} we rederive, from Poynting's theorem, the expression for the energy density obtained in Section \ref{sec:dipoles}, and show that it reduces to the result obtained in I in the special case of a homogeneous medium in which the complex permittivity has no spatial dependence. Section \ref{sec:stress} presents a brief derivation of the Casimir force density for a dispersive and dissipative dielectric based on the Maxwell stress tensor. Our conclusions are briefly summarized and discussed further in Section \ref{sec:conclusions}. As in I we focus mainly on the case of zero temperature. The extension to finite equilibrium temperatures is straightforward and so we simply summarize it in the Appendix, where we also briefly outline the derivation of the Lifshitz formula based on the force density (\ref{intro1}) or its finite-temperature generalization.


\section{Energy and Force Density from Coupled Dipoles}\label{sec:dipoles}
\subsection{Interaction Energy in Terms of Free-Space Green Dyadic}
We start by considering a collection of $\NN$ electric dipoles \emph{in free space}. The $n$th dipole (``atom") has associated with it an electric dipole moment operator $\bp_n=e\bx_n$ and a resonance frequency $\om_0$, and it is coupled to a reservoir that results in a damping rate $\gamma$ and a Langevin force $\bF_{Ln}$. It is also coupled to the total electric field $\bE(\br_n,t)$ at its position $\br_n$. The Heisenberg equation of motion for the electron coordinate operator for the $n$th atom is derived in I [Eq. (51)]\cite{fordkac}
\be
\ddot{\hx}_{ni}+\gamma\dot{\hx}_{ni}+\om_0^2\hx_{ni}=\frac{1}{m}\hat{F}_{Lni}(t)+\frac{e}{m}\hat{E}_i(\br_n,t).
\label{heq1}
\ee
The first subscript ($n$) on $\hx$ identifies the $n$th atom, while the second subscript ($i$) denotes 
the $i$th Cartesian component of $\bx_n$. As in I we use a circumflex to denote quantum-mechanical operators, and we define Fourier-transformed operators $\hp_{ni}(\om)=e\hx_{ni}(\om)$, etc. by writing
\be
\hp_{ni}(t)=\int_0^{\infty}d\om[\hp_{ni}(\om)e^{-i\om t}+\hp_{ni}(-\om)e^{i\om t}],
\ee
\be
\hF_{Lni}(t)=\int_0^{\infty}d\om[\hF_{Lni}(\om)e^{-i\om t}+\hF_{Lni}(-\om)e^{i\om t}],
\ee
\be
\hE_{i}(\br,t)=\int_0^{\infty}d\om[\hE_{i}(\br,\om)e^{-i\om t}+\hE_{i}(\br,-\om)e^{i\om t}],
\label{fourier}
\ee
with $\hp_{ni}(-\om)=\hp_{ni}^{\dag}(\om)$, etc., which follows from the requirement that the operators $\hp$, $\hF_L$, and $\hE$ be Hermitian. From (\ref{heq1}),
\be
[\om_0^2-\om^2-i\gamma\om]\hp_{ni}(\om)=\frac{e}{m}\hF_{Lni}(\om)+\frac{e^2}{m}\hE_i(\br_n,\om).
\label{heq2}
\ee

The electric field in (\ref{heq2}) is the sum of the source-free (``vacuum") field $\hE_{0i}(\br_n,\om)$ at $\br_n$ and the fields from all the dipole sources:
\be
\hE_i(\br_n,\om)=\hE_{0i}(\br_n,\om)+\frac{\om^2}{c^2}\sum_{m=1}^{\NN}G_{ij}^0(\br_n,\br_m,\om)\hp_{mj}(\om),
\ee
where ${G}^0(\br,\br',\om)$ is the Fourier transform of the (retarded) {\sl free-space} dyadic Green function
satisfying
\be
\nabla\times\nabla\times G^0({\bf r},{\bf r}')-\frac{\om^2}{c^2}G^0({\bf r},{\bf r}') =4\pi\delta^3(\br-\br').
\label{green0}
\ee
Equation (\ref{heq2}) is therefore
\bea
[\om_0^2-\om^2&-&i\gamma\om]\hp_{ni}(\om)=\frac{e}{m}\hF_{Lni}(\om)+\frac{e^2}{m}\hE_{0i}(\br_n,\om)\nonumber \\
&+&\frac{e^2}{m}\frac{\om^2}{c^2}\sum_{m=1}^{\NN}G_{ij}^0(\br_n,\br_m,\om)\hp_{mj}(\om), \nonumber \\
\label{couposc}
\eea
or, in matrix form,
\be
p(\om)=A(\om)\left[\frac{e}{m}F_L(\om)+\frac{e^2}{m}E_0(\om)\right],
\label{heq3}
\ee
where $p(\om)$, $F_L(\om)$, and $E_0(\om)$   are $3\NN$-dimensional column vectors ($\NN$ dipoles, $3$ Cartesian coordinates) and the $3\NN\times 3\NN$ matrix
\be
A(\om)=\left[\om_0^2-\om^2-i\gamma\om-\frac{e^2}{m}G_0\right]^{-1},
\label{pol1}
\ee
where
\be
G_0(\br,\br',\om)= \frac{\om^2}{c^2}G^0(\br,\br',\om).
\ee

We are interested in the change in energy involved in bringing the $\NN$ dipoles from a configuration where they are infinitely far apart and not interacting to one where they are separated by finite distances and interacting with themselves as well as with the source-free, fluctuating electric field. Consider first the simple example of a single dipole with real polarizability $\alpha(\om)$. According to the Hellmann-Feynman theorem \cite{hellfeyn} the change in energy when the dipole is brought from infinity (${\bf E}=0$) to a point $\br$ where the electric field of frequency $\om$ is ${\bf E}$ is 
\be
W=-\int_0^1\frac{d\lambda}{\lambda}\la\bp\cdot{\bf E}^{\dag}\ra=-\int_0^1\frac{d\lambda}{\lambda}\alpha(\om,\lambda)
\la{\bf E}(\br)\cdot{\bf E}^{\dag}(\br)\ra,
\label{hf1}
\ee
where the coupling constant in the integrand, in this case the electric charge, is taken to be $\lambda e$. Since
$\alpha(\om)$ is proportional to $e^2$, i.e., $\alpha(\om,\lambda)=\lambda^2\alpha(\om)$, Eq. (\ref{hf1}) simply generalizes the classical formula (\ref{intro2}). The relation \cite{greenident} [see also Eq. (\ref{greenident2}) below] between the vacuum expectation value of the free-field operator $\bE_0(\br,\om)\cdot\bE^{\dag}_0(\br,\om')$ and the Green dyadic $G_0$
\be
\la\bE_0(\br,\om)\cdot\bE^{\dag}_0(\br,\om')\ra=\frac{\hbar}{\pi}{\rm Im}{\rm Tr}\,G_{0}(\br,\br,\om)\delta(\om-\om'),
\label{hf2}
\ee
where ${\rm Tr}$ is the $3\times3$ trace, suggests that
\be
W=- \frac{\hbar}{\pi} \int_0^1\frac{d\lambda}{\lambda} \int_0^{\infty}d\om\alpha(\om,\lambda)
{\rm Im}{\rm Tr}\,G_{0}(\br,\br,\om)
\label{hf3}
\ee
when the field is the source-free electric field including all frequencies. But of course the polarizability
cannot be real at all frequencies, and the correct form of $W$ is not (\ref{hf3}) but  \cite{pwm3}
\be
W=-\frac{\hbar}{\pi}{\rm Im}\int_0^1\frac{d\lambda}{\lambda}\int_0^{\infty}d\om{\rm Tr}\alpha(\om,\lambda)
G_{0}(\br,\br,\om).
\label{hf4}
\ee
Thus if we again use $\alpha(\om,\lambda)=\lambda^2\alpha(\om)$, we recover the expression
\be
W=-\frac{\hbar}{2\pi}{\rm Im}\int_0^{\infty}d\om{\rm Tr}\alpha(\om)G_{0}(\br,\br,\om)
\label{hf5}
\ee
obtained by other methods \cite{pwm3}.

For our collection of dipoles we identify from (\ref{heq3}) the polarizability matrix $(e^2/m)A(\om)$ and define
the expectation value of the interaction energy as \cite{stark}
\bea
\cE&=&-\frac{\hbar}{\pi}{\rm Im}\int_0^{\infty}d\om{\rm Tr}\int_0^1\frac{d\lambda}{\lambda}\nonumber \\
&&\mbox{}\times\frac{\lambda^2(e^2/m)G_0}{\om_0^2-\om^2-i\gamma\om-\lambda^2(e^2/m)G_0}\nonumber \\
&=&\frac{\hbar}{2\pi}{\rm Im}\int_0^{\infty}d\om{\rm Tr}\log[1-\alpha_0G_0],
\label{heq4}
\eea
where we define the polarizability
\be
\alpha_0(\om)=\frac{e^2/m}{\om_0^2-\om^2-i\gamma\om}.
\label{defalpha}
\ee
(Spatial coordinates $\br_n,\br_m$ are treated here along with Cartesian components $i,j$ as matrix indices.)

The integrand in (\ref{heq4}) can be expanded in powers 
of $\alpha_0$:
\bea
{\cal E}&=&-\frac{\hbar}{2\pi}{\rm Im}\int_{0}^{\infty}d\om{\rm Tr}[\alpha_0G_0+\frac{1}{2}\alpha_0^2G_0G_0 
\nonumber \\
&&\mbox{}+\frac{1}{3}\alpha_0^3G_0G_0G_0+ ...].
\label{heq44}
\eea
The first term in brackets is part of a single-particle self-energy, while the terms that are non-diagonal in the
space coordinates $\br_n,\br_m$ correspond successively to two-body, three-body, etc. van der Waals interactions \cite{igor,mahan,renne,agarwal}. Thus, for instance, the second term in brackets, written out explicitly using
\bea
{\rm Tr}[\frac{1}{2}\alpha_0^2G_0G_0]&=&\frac{1}{2}\sum_{m,n=1}^{\NN}\alpha_0^2(\om)G_{0ij}(\br_n,\br_m,\om)\nonumber \\
&&\mbox{}\times G_{0ji}(\br_m,\br_n,\om),
\label{heq5}
\eea
is found to be just the sum of pairwise van der Waals interaction energies of the
$\NN$ atoms when terms with $m=n$ are excluded (these correspond to self-energies).  
In the model in which the atoms form a continuum we replace the
summation in (\ref{heq5}) by
\bea
&&\int d^3r\int d^3r' N(\br)N(\br')\alpha_0^2(\om) G_{0ij}(\br,\br',\om)G_{0ji}(\br',\br,\om)\nonumber \\
&& \equiv{\rm Tr}\left[\frac{\eps-1}{4\pi}\right]^2[G_0]^2,
\label{heq34}
\eea
where $N(\br)$ is the number of atoms per unit volume at $\br$. Using the formula 
\be
\eps(\om)=1+4\pi N\alpha_0(\om)=1+\frac{Ne^2/m}{\om_0^2-\om^2-i\gamma\om}
\label{perm}
\ee
for the permittivity, we similarly replace (\ref{heq44}) by
\be
{\cal E}=-\frac{\hbar}{2\pi}{\rm Im}\int_{0}^{\infty}d\om{\rm Tr}\sum_{n=1}^{\infty}
\frac{1}{n}\left[\frac{\eps-1}{4\pi}\right]^n\left[G_0\right]^n
\label{heq444}
\ee
and define the energy density
\be
u(\br)=\frac{\hbar}{2\pi}{\rm Im}{\rm Tr}\int_{0}^{\infty}d\om\log\left[1-\frac{\eps-1}{4\pi}G_0\right]
\label{heq6}
\ee
in the continuum approximation \cite{dimen}. We have presumed in deriving this expression that the polarizability $\alpha_0$
does not depend on $\br$, so that the dependence of the permittivity on $\br$ stems solely from the $\br$-dependence
of the number density $N(\br)$. However, it is straightforward to rederive (\ref{heq6}) with an $\br$-dependent 
$\alpha_0$, so that different parts of the medium can have different resonance frequencies as well as
different number densities.


\subsection{Energy Density in Terms of the Full Green Dyadic} \label{subsec:dipolesB}

$G^0$ satisfies (\ref{green0}), while the Green dyadic $G$ in the case of a medium with complex permittivity $\eps(\br,\om)$ satisfies 
\be
\nabla\times\nabla\times G({\bf r},{\bf r}')-\frac{\om^2}{c^2}G({\bf r},{\bf r}') =4\pi\delta^3(\br-\br'),
\label{green1}
\ee
together with appropriate boundary conditions. Therefore
\be
\nabla\times\nabla\times (G-G^0)-\frac{\om^2}{c^2}(G-G^0)=\frac{\om^2}{c^2}[\eps-1]G,
\ee
and the solution of this equation obtained using the Green function $G^0$ implies the Born-Dyson-type relation
\be
G=G^0+G^0\left[\frac{\om^2}{c^2}\frac{\eps-1}{4\pi}\right]G ,
\label{born}
\ee
i.e.,
\bea
G_{ij}(\br,\br,\om)&=&G_{ij}^0(\br,\br,\om)+\frac{\om^2}{c^2}\int d^3r'G_{ik}^0(\br,\br',\om)\nonumber \\
&&\mbox{}\times \frac{\eps(\br',\om)-1}{4\pi}G_{ki}(\br',\br,\om).
\label{ident}
\eea
From (\ref{heq6}) and (\ref{born}) it follows that
\be
{u}(\br)=- \frac{\hbar}{2\pi}{\rm Im} \int_0^{\infty}d\om {\rm Tr} \log[(G^0)^{-1} G] ,
\label{heq101}
\ee
which is the well-known  ``trace-log formula" \cite{tracelog}. 
The free-space Green dyadic $G^0$ is independent of the atoms or the properties of the medium formed by them.
The term $\log[(G^0)^{-1}]$ in the above expression therefore subtracts from the total Green function $G$ the ``bulk" contribution
$G^0$, and therefore $u(\br)$ contains only the ``scattering" part of the Green function, that comprises the interaction between the atoms.

We can express this energy in a different form by partial integration as in Reference \cite{milton}. Omitting
the ``bulk" contribution in (\ref{heq101}),
\bea
{u}(\br)&=&\frac{\hbar}{2\pi}{\rm Im}{\rm Tr}\int_0^{\infty}d\om\om\frac{\pa}{\pa\om}\log G \nonumber \\
&=&\frac{\hbar}{2\pi}{\rm Im}{\rm Tr}\int_0^{\infty}d\om\om G^{-1}G' ,
\label{milt1}
\eea
where we use a prime to denote differentiation with respect to $\om$. Now from (\ref{green1}),
\be
\nabla\times\nabla\times \delta G-\frac{\om^2}{c^2}\eps\delta G=\frac{\om^2}{c^2}\delta\eps G,
\ee
and therefore, formally,
\be
\delta G=\frac{1}{4\pi}\frac{\om^2}{c^2}\delta\eps GG,
\label{thfour}
\ee
where of course there is an integration over space implied on the right-hand side. Similarly
\be
\nabla\times\nabla\times G'-\frac{\om^2}{c^2}\eps G'=\frac{2\om}{c^2}\eps G+\frac{\om^2}{c^2}\eps'G,
\label{id1}
\ee
and therefore
\be
G'=\frac{1}{4\pi}\frac{\om}{c^2}[2\eps+\om\eps']GG .
\label{id2}
\ee
Using this result in
(\ref{milt1}), we obtain
\bea
{u}(\br) &=&\frac{\hbar}{8\pi^2c^2}{\rm Im}{\rm Tr}\int_0^{\infty}d\om\om^2[2\eps(\br,\om)+\om\eps'(\br,\om)]
\nonumber \\
&&\mbox{}\times G(\br,\br,\om).
\label{enn}
\eea
This is our general expression for the energy density. We discuss it further and give an alternative derivation of it in Section \ref{sec:macro}. 

Two points concerning the derivation of (\ref{enn}) are worth noting here. First, we have seen that the energy 
density (\ref{heq6}) is associated with many-body van der Waals interactions, and therefore vanishes when the polarizability $\alpha_0(\om)\rightarrow 0$ and the permittivity $\eps(\om)\rightarrow 1$ for all frequencies, as can also be seen from (\ref{heq101}). However, when $\eps(\om)\rightarrow 1$ the energy density (\ref{enn}) that was derived 
from (\ref{heq6}) becomes
\bea
{u}(\br)&=&\frac{\hbar}{4\pi^2c^2}{\rm Tr}\int_0^{\infty}d\om\om^2 {\rm Im}G^0(\br,\br,\om) \nonumber \\
&=&2\int_0^{\infty}d\om\Big(\frac{1}{2}\hbar\om\Big)\Big(\frac{\om^2}{2 \pi^2c^3}\Big),
\label{enn2}
\eea
where we have used the fact that ${\rm Tr}{\rm Im}G^0(\br,\br,\om)=2\om/c$. Thus we recover the correct vacuum electromagnetic energy density in the limit $\eps(\om)\rightarrow 1$. The reason for this is that in obtaining the trace-log formula (\ref{milt1}) from (\ref{heq101}) we have effectively added the part of the energy density needed to make $u(\br)$ have the correct (nonvanishing) vacuum-field energy density when $\eps(\om)\rightarrow 1$.

The second point is that there are other contributions to the total energy density that are not
included in (\ref{enn}) and that do not contribute to the Casimir force density. Of course this is not surprising, as in deriving (\ref{enn}) we considered only the energy involved in the induction of dipole moments by the electric field and in the interaction of these dipoles. Thus our derivation does not account for the constant energy (density) absorption rate of a medium with permittivity $\eps(\br,\om)$; this absorption rate $R_{\rm abs}$ has exactly the form expected from classical electromagnetic theory:
\bea
R_{\rm abs}&=&\frac{1}{4\pi}{\rm Tr}\int_0^{\infty}d\om\om\eps_I(\br,\om)\la\bE(\br,\om)\cdot\bE^{\dag}(\br,\om)\ra,\nonumber \\
&=&\frac{\hbar}{4\pi^2c^2}{\rm Tr}\int_0^{\infty}d\om\om^3\eps_I(\br,\om){\rm Im}G(\br,\br,\om),
\label{rabs}
\eea
where in the second line we have used the generalization of (\ref{hf2}) given in  (\ref{greenident2}) below. This flow of power into the atoms' reservoirs is of course exactly cancelled by the power lost by the field, so that there is no net change in total energy density. In the calculation using the Poynting theorem in the following section, all contributions to the total energy density are included, and this cancellation is seen explicitly.

In addition to the fluctuating dipoles induced by the electromagnetic field, whose interaction is responsible for van der Waals (Casimir) forces, there are fluctuating dipoles due to the
Langevin noise forces acting on the individual atoms. These ``noise" dipole moments do not result in forces among the atoms, since the Langevin forces acting on different atoms, unlike the electric fields inducing dipole moments in different atoms, have no spatial correlations. In the presence of the fluctuating electric field they do, however, contribute 
($\hat{\bf K} = 4 \pi \hat{\bf P}_{\rm noise}$ is the noise polarization defined in Eq. (68) of I)
\begin{widetext}
\bea
u_N(\br)&=&\frac{1}{8\pi}\la\bE\cdot{\hat{\bf D}}_{\rm noise}\ra=\frac{1}{2}\la\bE\cdot{\hat{\bf P}}_{\rm noise}\ra \nonumber \\
&=&\frac{1}{2}{\rm Re}\int_0^{\infty}d\om\int_0^{\infty}d\om'\la\hat{\bf P}_{\rm noise}(\br,\om)\cdot
\hat{\bf E}^{\dagger}(\br,\om')\ra e^{-i(\omega-\omega')t}
=\frac{1}{8\pi}{\rm Re}\int_0^{\infty}d\om\int_0^{\infty}d\om'\la\hat{\bf K}(\br,\om)\cdot\hat{\bf E}^{\dagger} (\br,\om')\ra e^{-i(\omega-\omega')t} \nonumber \\
&=&\frac{1}{8\pi}{\rm Re}\int_0^{\infty}d\om\int_0^{\infty}d\om'\frac{\om^2}{4\pi c^2}G_{ij}(\br,\br,\om') \; \la\hat{K}_i(\br,\om)\hat{K}_j^{\dagger}(\br,\om')\ra 
e^{-i(\omega-\omega')t} \nonumber \\
&=&\frac{\hbar}{8\pi^2c^2}\int_0^{\infty}d\om\om^2\eps_I(\om)G_{Rii}(\br,\br,\om)
\label{enoise}
\eea
\end{widetext}
to the total energy density. 
This is a single-particle self-energy resulting in effect from
the interaction of the Langevin-force-induced dipole moment of each particle with the part of 
the electric field due to  this same dipole moment. As such it has the effect in our model 
of determining in part the equilibrium positions of the particles within the medium, 
but does not contribute to interparticle interactions.


\subsection{Force Density}
The force density can be obtained from either expression (\ref{heq6}) or (\ref{enn}) for the energy density. Consider first 
(\ref{heq6}). The fact that $G^0$ is independent of any properties of the medium means that a variation $\delta{\cal E}$ of ${\cal E}$ due to a deformation of the medium depends only on the variation $\delta\eps$ of $\eps$ that accompanies the deformation \cite{igor}. Therefore
\bea
\delta{\cal E}&=&-\frac{\hbar}{2\pi}{\rm Im}\int_{0}^{\infty}d\om{\rm Tr}\frac{\delta\eps}{4\pi}\sum_{n=0}^{\infty}
\left[\frac{\eps-1}{4\pi}\right]^n\left[G_0\right]^{n+1}\nonumber \\
&=&-\frac{\hbar}{8\pi^2}{\rm Im}\int_{0}^{\infty}d\om\int d^3r\delta\eps(\br,\om)
\Big\{G_{0ii}(\br,\br,\om)\nonumber \\
&&\mbox{}+\int d^3r'G_{0ik}(\br,\br',\om)\left[\frac{\eps(\br',\om)-1}{4\pi}\right]G_{0ki}(\br',\br,\om)\nonumber \\
&&\mbox{}+\int d^3r'\int d^3r''G_{0ik}(\br,\br',\om)\left[\frac{\eps(\br',\om)-1}{4\pi}\right]
\nonumber \\
&&\mbox{}\times G_{0kp}(\br',\br'',\om)\left[\frac{\eps(\br'',\om)-1}{4\pi}\right] 
G_{0pi}(\br'',\br,\om)\nonumber \\
&&\mbox{}+ ...\Big\}.
\label{heq7}
\eea 
It follows from (\ref{ident}) that the quantity in curly brackets is just the trace of the Green dyadic $G(\br,\br,\om)$, and consequently the variation in the energy density is
\be
\delta{u}(\br)=-\frac{\hbar}{8\pi^2c^2}{\rm Im}\int_{0}^{\infty}d\om\om^2\delta\eps(\br,\om){\rm Tr}\,G(\br,\br,\om).
\ee

Following Dzyaloshinskii \emph{et al.} \cite{igor}, we consider an infinitesimal local displacement ${\bf R}(\br)$ of the medium. This transport of the medium implies a variation in $\eps$ at $\br$ such that $\eps(\br,\om)+\delta\eps(\br,\om)=\eps(\br-{\bf R},\om)$, or $\delta\eps(\br,\om)=-\nabla\eps\cdot{\bf R}$. Therefore
\bea
\delta{u}(\br)&=&-\frac{\hbar}{8\pi^2c^2}{\rm Im}\int_{0}^{\infty}d\om\om^2[-\nabla\eps\cdot{\bf R}]{\rm Tr}\,G(\br,\br,\om)
\nonumber \\
&\equiv&-\int d^3r{\bf f}(\br)\cdot{\bf R},
\label{heq100}
\eea
where the Casimir force density ${\bf f}(\br)$ obtained in this way is given by equation (\ref{intro1}): 
\be
{\bf f}(\br)=-\frac{\hbar}{8\pi^2c^2}{\rm Im}\int_{0}^{\infty}d\om\om^2\nabla\eps(\br,\om)G_{ii}(\br,\br,\om),
\label{intro1again}
\ee
where we used that ${\rm Tr}\,G \equiv G_{ii}$. This is equivalent to the result of Dzyaloshinskii \emph{et al.} 
obtained by summing a sequence of diagrams, each successive one including a number $n$ of closed
loops corresponding to $n$-body van der Waals interaction energies.
As these authors discuss, equation (\ref{intro1}) does not in general give the total force density, as one must
account for the variation in $\eps$ due to changes in density as well as to the displacements considered in obtaining
(\ref{heq100}). Ignoring the former amounts to assuming displacements ${\bf R}$ such that $\nabla\cdot{\bf R}=0$.
For the calculation of forces on bodies embedded in uniform fluids in mechanical equilibrium, however, the
additional contribution to the total force resulting from density variations must be balanced by pressure arising
from non-electromagnetic forces, and may in effect be omitted \cite{igor}. It must also be noted that $G(\br,\br,\om)$
diverges unless spatial frequencies are cut off at some large value on the order of 1/(interatomic spacing).
However, again as discussed by Dzyaloshinskii \emph{et al.}, the contributions from high spatial frequencies
are the same at each point $\br$ for an inhomogeneous medium as for a homogeneous medium having the same value of $\eps$
at $\br$ as the inhomogeneous medium. The divergent Green dyadic appearing in (\ref{intro1}) should therefore be
replaced by \cite{igor} 
\be
G(\br,\br,\om)-\overline{G}(\br,\br,\om),
\ee
where $\overline{G}$ is the Green function of a homogeneous medium with the same value of $\eps$ at $\br$ as the
inhomogeneous medium under consideration \cite{tip}. This subtraction of the singular $\overline{G}(\br,\br,\om)$ is discussed further below. 

Consider now the derivation of the force density from the energy density as expressed by (\ref{enn}).
A variation $\delta\eps$ in the permittivity results in a variation
\bea
\delta u(\br)&=&\frac{\hbar}{8\pi^2c^2}{\rm Im}\int_0^{\infty}d\om\om^2\Big\{[2\delta\eps+\om\delta\eps']G_{ii}
\nonumber \\
&&\mbox{}+[2\eps+\om\eps']\delta G_{ii}\Big\}
\label{var1}
\eea
in $u(\br)$. From (\ref{thfour}) and (\ref{id2}),
\bea
[2\delta\eps&+&\om\delta\eps']G+[2\eps+\om\eps']\delta G=[2\delta\eps+\om\delta\eps']G\nonumber \\
&&\mbox{}+[2\eps+\om\eps']\frac{1}{4\pi}\frac{\om^2}{c^2}\delta\eps GG 
=[2\delta\eps+\om\delta\eps']G+\om\delta\eps G'\nonumber \\
&=&2\delta\eps G +\om\frac{\pa}{\pa\om}[\delta\eps G].
\eea
Partial integration of (\ref{var1}) then gives
\be
\delta u(\br)=-\frac{\hbar}{8\pi^2c^2}{\rm Im}\int_0^{\infty}d\om\om^2\delta\eps(\br,\om)G_{ii}(\br,\br,\om),
\ee
implying again the force density (\ref{intro1}) when $\eps$ varies with position. Essentially the same calculation was presented by Milton {\sl et al.} \cite{milton} under the assumption that $\eps_I=0$.

An ``inhomogeneous medium" as defined here includes the important case of spatially separated homogeneous
bodies, as in the case of the two parallel dielectric slabs treated by Lifshitz \cite{lifshitz}. Except for
such simple geometries, the calculation of Casimir forces based on (\ref{intro1})---which amounts in effect to
the calculation of the (classical) dyadic Green function---must be performed numerically. Our interest
here, as discussed in the Introduction, is only in the general question regarding what effect dissipation has on the derivation of (\ref{intro1}) by different methods, especially in derivations based on energy variations.


\subsection{Remarks Relating to Early Work} \label{susect2d}
In their diagrammatic approach Dzyaloshinskii \emph{et al.} obtained the following expression for the force
density (in addition to the aforementioned contribution that arises from displacements ${\bf R}$ with $\nabla\cdot{\bf R}\neq 0$): 
\be
{\bf f}(\br)=-\frac{k_BT}{4\pi c^2} \sum_{n=0}^{\infty'}  \om_n^2G_{ii}(\br,\br,i\om_n)\nabla\eps(\br,i\om_n).
\label{dd1}
\ee
Here the (Matsubara) frequencies $\om_n=2\pi nk_BT/\hbar$, and the prime 
on the zero below the summation sign means that a factor $1/2$
multiplies the $n=0$ term. For $T\rightarrow 0$ the sum can be replaced by an integral, and
(\ref{dd1}) becomes
\be
{\bf f}(\br)=-\frac{\hbar}{8\pi^2c^2}\int_0^{\infty}d\xi\xi^2\nabla\eps(\br,i\xi)G_{ii}(\br,\br,i\xi).
\label{heq51}
\ee
$G$, like $\eps$, is analytic in the upper half of the complex frequency plane. Using this fact, it is
easy to show that (\ref{heq51}) is equivalent to (\ref{intro1}). The
appearance of the permittivity at imaginary frequencies in the work of Dzyaloshinskii \emph{et al.} can be traced
to an analytic continuation of a Green function obtained with $\eps$ evaluated at real frequencies to one
in which $\eps$ is evaluated at imaginary frequencies. Thus, consistent with the remarks in the Introduction about
$\eps$ being a function that is real-valued on the imaginary axis, the validity of the derived force density for dissipative media rests on the formal analytic properties of $\eps$. The same is true of the analysis of 
Schwinger \emph{et al.} \cite{julie}. Our more ``brute-force" approach does not explicitly invoke these properties; we calculate the Casimir free energy by exhibiting explicitly, albeit with a simple model, the dissipative character of the dielectric. 

Mahan \cite{mahan} considered essentially equation (\ref{heq1}) without damping and Langevin forces and without
coupling to the fluctuating source-free field. The eigenfrequencies $\Om_j$ of the coupled-oscillator system are
obtained by writing ${\bf x}_n={\bf x}_n(\Om)\exp(-i\Om t)$ and then applying the argument theorem to obtain the
difference in the zero-point energies $\sum_j(1/2)\hbar\Om_j$ between a given configuration and that when all the
interatomic distances are infinite. Renne \cite{renne} noted inconsistencies in this approach in the case of retarded
interactions, and proceeded in a similar fashion by solving for the eigenfrequencies of the coupled system
of dipole oscillators and field oscillators. 

Renne includes radiative reaction, as does Agarwal \cite{agarwal}
in a general response-function formulation, and consequently the polarizabilities
of the material oscillators are complex; radiative reaction makes the only contribution to the imaginary part of
the polarizability in their work. However, this imaginary part is connected not with dissipation of field energy as such but with ({\sl energy-conserving}) Rayleigh scattering of radiation, consistent with the optical theorem. To compare with the result of Renne, for instance, for the interaction energy of a system of ground-state harmonic dipole oscillators, we use the identity ${\rm Tr}\log X=\log{\rm det}X$ \cite{julians} to write (\ref{heq4}) as
\bea
{\cal E}&=&\frac{\hbar}{2\pi}{\rm Im}\int_0^{\infty}d\om\log{\rm det}[1-\alpha_0(\om)G_0(\om)]\nonumber \\
&=&\frac{\hbar}{2\pi}\int_0^{\infty} d\xi\log{\rm det}[1-\alpha_0(i\xi)G_0(i\xi)],
\eea
which is equivalent to equation (18) of Renne \cite{renne}. In his derivation, however, the damping rate $\gamma$ in the
definition (\ref{defalpha}) of $\alpha_0(\om)$ is obtained from an {\sl approximation} to the radiative reaction
field in which the third derivation with respect to time of a dipole moment is replaced by minus the square of an oscillation frequency times a first derivative with respect to time, and in this approximation the imaginary part of $\alpha_0(\om)$ provides in effect for a collisional- or ohmic-type dissipation rather than for elastic scattering of radiation. (Without this approximation the permittivity in Renne's approach would not be analytic in the upper half of the complex frequency plane, and it would preclude the analytic continuation of
a Green function that was alluded to earlier.) In our approach $\gamma$ is obtained, without approximation, from the coupling of each oscillator to its reservoir, and $\eps(\om)$ [Eq. (\ref{perm})] has no contribution from radiative reaction. Dissipation of field energy is due, as usual, simply to the transfer of energy from the field to the atoms and the subsequent transfer of this energy to the reservoirs rather than back to the field. 

Radiative reaction is of course naturally included in our approach; it is associated with the term $G^0_{ij}(\br_n,\br_n,\om)$ in equation (\ref{couposc}) and with a renormalizable self-energy. The complex 
permittivity (\ref{perm}) appearing in equation (\ref{green1}) for the Green function, however, has no contribution from radiative reaction. This is related to the implicit assumption in our
continuum model that there are no density fluctuations and therefore that the extinction coefficient due to Rayleigh scattering vanishes \cite{rosenfeld}. In the case of two spatially separated homogeneous bodies, for example, the only ``scattering" that occurs is in the form of reflection and refraction at boundaries.


\section{Macroscopic QED Approach to the Energy Density of a Dissipative Dielectric Medium}\label{sec:macro}
In I we derived expressions for quantized electric and magnetic fields and for the energy density
in a dispersive and absorbing, homogeneous dielectric medium. We now extend these considerations
to inhomogeneous media.


\subsection{Poynting's Theorem}
As in I we base our approach essentially on the macroscopic, Heisenberg-picture Maxwell
equations and the Poynting's theorem that follows from them.
In the conventional notation Poynting's theorem for the symmetrized Poynting vector operator
$\hat{\bf S}=(c/8\pi)[\bE\times\bH-\bH\times\bE]$ takes the form
\bea
\label{Poynting3}
\oint_S\la\hat{\bf S}\ra\cdot{\bf n}\,da&=&-{1\over 8\pi}\int_V\la\bE\cdot{\pa\bD\over\pa t}+{\pa\bD\over\pa t}
\cdot\bE\ra d^3r
\nonumber \\
\mbox{}&&-{1\over 8\pi}\int_V\la\bH\cdot{\pa\bB\over\pa t}+{\pa\bB\over\pa t}\cdot\bH\ra d^3r \nonumber \\
&=&-\int_V \frac{\pa}{\pa t}U(\br,t) d^3r, 
\eea
or, after a time-integration,
\bea
\label{Poynting3TI}
\int_0^t dt' \oint_S\la\hat{\bf S}\ra\cdot{\bf n}\,da = - \int_V \left[ {U}({\bf r},t) - {U}({\bf r},-\infty) \right] d^3r, 
\eea
where the expectation values refer to the ground state of the system consisting of the field, the atoms, 
and the reservoir oscillators responsible for the damping and Langevin forces acting on the atoms, and
also
\begin{eqnarray}
\label{EnDensity} 
{U}({\bf r},t) - {U}({\bf r},-\infty) &=& \frac{1}{8\pi} \int_{-\infty}^{t}dt' 
\langle \hat{\bf E} \cdot \frac{\pa\hat{\bf D}}{\pa t'} + \frac{\pa\hat{\bf D}}{\pa t'}\cdot \hat{\bf E}  \nonumber \\  
&&\mbox{} +  \hat{\bf H} \cdot \frac{\pa \hat{\bf B}}{\pa t'} + \frac{\pa \hat{\bf B}}{\pa t'} \cdot \hat{\bf H}\rangle.
\end{eqnarray}
The left-hand side of (\ref{Poynting3}) is the electromagnetic energy flux through a surface $S$ 
enclosing the volume $V$, meaning that ${U}({\bf r},t) - {U}({\bf r},-\infty)$ is the energy density variation
associated to this flux (through $S$) between $t'=-\infty$ and $t'=t$. The latter is immediately seen to vanish for static systems in equilibrium, but this can be circumvented by imagining that we bring 
our system adiabatically from some reference configuration \cite{BarashGinzburg}. We can then
use the approximation that the system is stationary at both $t'=-\infty$ and $t'=t$, but as those instants are
now characterized by different configurations, the subtraction in (\ref{EnDensity}) does not vanish anymore.
In addition, by choosing our reference configuration as one where the different parts of the system do not
interact, we can just subtract ${U}({\bf r},-\infty)$ 
from (\ref{EnDensity}) (as an irrelevant constant) and write
\begin{eqnarray}
\label{EnDensity2} 
{U}({\bf r},t) = \frac{1}{8\pi} \int_{-\infty}^{t} \!\! dt' && \hspace{-12pt} 
\langle \hat{\bf E} \cdot \frac{\partial\hat{\bf D}}{\pa t'} + \frac{\partial \hat{\bf D}}{\pa t'} \cdot \hat{\bf E}  \nonumber \\  
&&\mbox{} +  \hat{\bf H} \cdot \frac{\partial \hat{\bf B}}{\pa t'} + \frac{\partial\hat{\bf B}}{\pa t'} \cdot \hat{\bf H}\rangle.
\label{poynt100}
\end{eqnarray}
We remark that the energy density $U({\bf r}, t)$, being a quantity that is associated
with the electromagnetic energy flux, is not necessarily the same as the energy density $u({\bf r})$
of the previous section; in particular, as we shall see, it contains the energy of the bath field.
In terms of the Fourier components of the fields [cf. (\ref{fourier})], the total equilibrium energy density is
\begin{eqnarray}
\label{EnDensityOm}
&& U({\bf r},t) = \frac{1}{8\pi} \int_{-\infty}^{\infty} \!\! d\omega \int_{-\infty}^{\infty}d\omega' \, \frac{e^{-i(\omega'+\omega)t}}{\omega'+\omega}
\nonumber \\
&&\times \left[ \omega' \langle \hat{\bf E}({\bf r},\omega) \cdot \hat{\bf D}({\bf r},\omega') \rangle + \omega \langle \hat{\bf D}({\bf r},\omega) \cdot \hat{\bf E}({\bf r},\omega') \rangle  \right. \nonumber \\
&&\hspace{-12pt}+ \left. \omega'\langle \hat{\bf H}({\bf r},\omega) \cdot \hat{\bf B}({\bf r},\omega') \rangle +  \omega \langle \hat{\bf B}({\bf r},\omega) \cdot \hat{\bf H}({\bf r},\omega') \rangle\right] \! . \;
\end{eqnarray}
For zero temperature we can write this in terms of strictly positive frequencies as follows: 
\be
\label{EnDensityOm2}
U({\bf r},t) =U_e({\bf r},t) + 
U_m({\bf r},t), 
\ee
where 
\bea
U_e({\bf r},t) &=&\frac{1}{8\pi}  \!\! \int_{0}^{\infty}d\omega \int_{0}^{\infty}d\omega' \, 
\frac{e^{i(\om'-\om)t}}{\omega'-\omega} \nonumber \\
&&\mbox{}\times \Big[ \om'\langle \hat{\bf E}({\bf r},\omega) \cdot
\hat{\bf D}^{\dagger}({\bf r},\omega') \rangle \nonumber \\
&&\mbox{}-\om\langle \hat{\bf D}({\bf r},\omega) 
\cdot \hat{\bf E}^{\dagger}({\bf r},\omega') \rangle \Big]  
\eea
and, for the non-magnetic media under consideration,
\bea
\label{U_m}
U_m({\bf r},t)&=& \frac{1}{8\pi} \int_{0}^{\infty} d\omega\la \hat{\bf H}(\br,\om)\cdot\hat{\bf H}^{\dagger}
(\br,\om)\ra.
\label{umag}
\eea
We have used the fact that $\hat{\bf E}(\br,-\om)=\hat{\bf E}^{\dagger}(\br,\om)$, etc. and, for $\omega, \omega' > 0$,
\begin{eqnarray}
\langle \hat{\bf E}({\bf r},\omega)\cdot \hat{\bf D}({\bf r},\omega') \rangle = \langle \hat{\bf E}^{\dagger}({\bf r},\omega)\cdot \hat{\bf D}^{\dagger}({\bf r},\omega') \rangle = 0,  
\end{eqnarray} 
and likewise for the corresponding bilinear magnetic field products. We have also made use of the fact that, for zero temperature, only the positive-frequency field operators $\hat{\bf E}^{\dagger}(\br,\om)$, etc. produce a nonvanishing
result when acting on the initial state of the system, i.e., they act as creation operators, whereas the
positive-frequency operators $\hat{\bf E}(\br,\om)$ are annihilation operators and yield zero when acting
on the initial zero-temperature state. 
\bigskip


\subsection{Electric and Magnetic Fields} \label{subsec:ElMag}
The electric field operator $\hat{\bf E}(\br,\om)$ satisfies 
\begin{eqnarray}
\label{EquationE}
[\nabla\times\nabla\times -\frac{\om^2}{c^2}\epsilon(\br,\om)]\hat{\bf E}(\br,\om)=\frac{\omega^2}{c^2} 
\hat{\bf K}({\bf r},\omega) .
\end{eqnarray}
The solution for $\hat{\bf E}(\br,\om)$ in the medium can be written in terms of the 
Green dyadic satisfying (\ref{green1}):
\begin{eqnarray}
\label{GreenEquation}
\hat{E}_i(\br,\om)=\frac{\om^2}{4\pi c^2}\int d^3r'G_{ij}(\br,\br',\om)\hat{K}_j(\br',\om),
\label{green101}
\end{eqnarray}
while for $\hat{\bf D}(\br,\om)$ we have
\be
\hat{\bf D}(\br,\om)=\eps(\br,\om)\hat{\bf E}(\br,\om)+\hat{\bf K}(\br,\om).
\label{Deq}
\ee
Based on the harmonic-oscillator reservoir model, it is shown in I, for instance, that the noise polarization $\hat{\bf K}$ has the thermal equilibrium properties
\bea 
&&\la \hK_i(\br,\om)\ra = \la \hK^{\dag}_i(\br,\om)\ra = 0 , \nonumber \\
&&\hspace{-15pt}\la \hK_i(\br,\om)\hK_j(\br',\om')\ra = \la \hK^{\dag}_i(\br,\om)\hK^{\dag}_j(\br',\om')\ra = 0 ,
\eea 
and
\bea
\la\hK^{\dag}_i(\br,\om)\hK_j(\br',\om')\ra&=&4\hbar\eps_I(\om)\delta_{ij}\delta(\om-\om')\delta^3(\br-\br')
\nonumber \\
&&\mbox{}\times{1\over e^{\hbar\om/k_BT}-1},
\label{prop1}
\eea
\bea
\la\hK_i(\br,\om)\hK^{\dag}_j(\br',\om')\ra&=&4\hbar\eps_I(\om)\delta_{ij}\delta(\om-\om')\delta^3(\br-\br')
\nonumber \\
&&\mbox{}\times \left[ {1\over e^{\hbar\om/k_BT}-1}+1 \right].
\label{prop2}
\eea
The expression for $\hat{\bf H}(\br,\om)$ in the medium follows from the Maxwell equation $\nabla\times\hat{\bf E}=-(1/c)\pa\hat{\bf H}/\pa t$:
\bea
\hat{\bf H}(\br,\om)&=&-i\frac{c}{\om}\nabla\times\hat{\bf E}(\br,\om)\nonumber \\
&=&-i\frac{c}{\om}\int d^3r'[\nabla\times G(\br,\br',\om)]\cdot\hat{\bf K}(\br',\om).\nonumber \\
\eea


\subsection{Total Energy Density} \label{subsec:TED}

Consider first the energy density $U_e$. The expression (\ref{Deq}) for $\hat{\bf D}(\br,\om)$ results in two
contributions to $U_e$. The contribution from $\eps(\br,\om)\hat{\bf E}(\br,\om)$ is
\bea
 U^{(1)}_e(\br,t)&=&\frac{1}{8\pi}{\rm Re}\int_0^{\infty}d\om\int_0^{\infty}d\om'\frac{e^{i(\om'-\om)t}}{\om'-\om}
\nonumber \\
&&\mbox{}\times\Big[\om'\eps^*(\om')-\om\eps(\om)\Big]\langle\hat{\bf E}(\br,\om)\cdot
\hat{\bf E}^{\dagger}(\br,\om')\rangle\nonumber \\
&=&\frac{\hbar}{8\pi^2c^2}{\rm Re}{\rm Tr}\int_0^{\infty}d\om\om^2\int_0^{\infty}d\om'\frac{e^{i(\om'-\om)t}}{\om'-\om}
\nonumber \\
&&\mbox{}\times\Big[\om'\eps^*(\om')-\om\eps(\om)\Big]G_I(\br,\br,\om)\delta(\om-\om'),\nonumber \\
\label{zpe22}
\eea
where to simplify notation we have suppressed the $\br$ dependence of $\eps$. We have employed the 
identity \cite{greenident} [see also Eq. (\ref{greenident2}) below]
\be
\langle\hat{\bf E}(\br,\om)\cdot\hat{\bf E}(\br,\om')\rangle=\frac{\hbar}{\pi}\frac{\om^2}{c^2}{\rm Tr}\,G_I(\br,\br,\om)
\delta(\om-\om'),
\label{green103}
\ee
with $G_I$ denoting the imaginary part of the Green dyadic.
In order to evaluate  (\ref{zpe22}) we write the permittivity in terms of its real and imaginary parts, 
$\eps(\om)=\eps_R(\om)+i\eps_I(\om)$). In the term containing $\eps_R(\om)$ we use
\begin{equation}
\label{Derivative}
\lim_{\omega' \rightarrow \omega} \frac{f(\omega') - f(\omega)}{\omega' - \omega}=\frac{d f(\omega)}{d \omega}.
\end{equation}
with $f(\om)=\om\eps_R(\om)$, while in the term containing $\eps_I(\om)$ we use  
\begin{equation}
\label{LimitExp}
\lim_{\omega' \rightarrow \omega} \frac{e^{i(\om'-\om)t} - e^{-i(\om'-\om)t}}{\om'-\om} = 2it .
\end{equation}
After straightforward manipulations we obtain 
\bea \label{U1}
 U^{(1)}_e(\br)&=&\frac{\hbar}{8\pi^2c^2}{\rm Tr}\int_0^{\infty}d\om\om^2\frac{\pa}{\pa\om}[\om\eps_R(\om)]G_I(\br,\br,\om)
\nonumber \\
&&\mbox +t\frac{\hbar}{4\pi^2c^2}{\rm Tr}\int_0^{\infty}d\om\om^3\eps_I(\om)G_I(\br,\br,\om).
\label{ue1}
\eea

The contribution to $U_e(\br,t)$ from the noise polarization part of $\hat{\bf D}$ is
\begin{widetext}
\bea
U_e^{(2)}(\br,t)&=&{\rm Re}\frac{1}{8\pi}\int_0^{\infty}d\om\int_0^{\infty}d\om'\frac{e^{i(\om'-\om)t}}{\om'-\om}
\Big[\om'\la\hat{\bf E}(\br,\om)\cdot\hat{\bf K}^{\dagger}(\br,\om')\ra-
\om\la\hat{\bf K}(\br,\om)\cdot\hat{\bf E}^{\dagger}(\br,\om')\ra\Big]\nonumber \\
&=&{\rm Re}\frac{1}{8\pi}\int_0^{\infty}d\om\int_0^{\infty}d\om'\frac{e^{i(\om'-\om)t}}{\om'-\om}
\frac{1}{4\pi}\Big[\frac{\om'\om^2}{c^2}\int d^3r'G_{ij}(\br,\br',\om)\la K_j(\br,\om)K_i^{\dagger}(\br',\om')\ra \nonumber \\
&&\mbox{}-\frac{\om\om'^2}{c^2}\int d^3r'G^*_{ij}(\br,\br',\om')\la K_i(\br,\om)K_j^{\dagger}(\br',\om')\ra\Big].
\label{zpe2}
\eea
\end{widetext}
Using (\ref{prop2}) with $T=0$, and the same sort of manipulations used to evaluate $U_e^{(1)}(\br)$, 
we obtain
\bea
U_e^{(2)}(\br,t)&=&-\frac{\hbar}{8\pi^2c^2}\int_0^{\infty}d\om\om^2\eps_I(\om)\frac{\pa}{\pa\om}[\om {\rm Tr}\,G_{R}(\br,\br,\om)]\nonumber \\
&&\mbox{}-t\frac{\hbar}{4\pi^2c^2}\int_0^{\infty}d\om\om^3\eps_I(\om){\rm Tr}\,G_{I}(\br,\br,\om).
\label{ue2}
\eea
Thus the terms proportional to $t$ in (\ref{ue1}) and (\ref{ue2}) cancel, as discussed in I for the special case of a homogeneous medium, and then
\bea \label{cancelation}
U_e(\br)&=&\frac{\hbar}{8\pi^2c^2}\int_0^{\infty}d\om\om^2\Big\{\frac{\pa}{\pa\om}[\om\eps_R(\om)]
{\rm Tr}\,G_{I}(\br,\br,\om)
\nonumber \\
&&\mbox{}-\eps_I(\om)\frac{\pa}{\pa\om}[\om {\rm Tr}\,G_{R}(\br,\br,\om)]\Big\}.
\eea
Integrating by parts the integral involving the second term on the right, we obtain
\bea
U_e(\br)&=&\frac{\hbar}{8\pi^2c^2}{\rm Im}\int_0^{\infty}d\om\om^2\frac{\pa}{\pa\om}[\om\eps(\om)]
{\rm Tr}\,G(\br,\br,\om)\nonumber \\
&&\mbox{}+\frac{\hbar}{8\pi^2c^2}\int_0^{\infty}d\om\om^2\eps_{I}(\om){\rm Tr}\,G_{R}(\br,\br,\om).
\eea
It follows similarly from (\ref{umag}) that
\be
U_m(\br)=\frac{\hbar}{8\pi^2c^2}{\rm Im}\int_0^{\infty}d\om\om^2\eps(\om){\rm Tr}\,G(\br,\br,\om).
\ee
The total zero-point energy density of the dispersive and dissipative dielectric medium is therefore
\bea
U(\br)&=&\frac{\hbar}{8\pi^2c^2}{\rm Im}\int_0^{\infty}d\om\om^2\Big\{\frac{\pa}{\pa\om}[\om\eps(\om)]+\eps(\om)\Big\}\nonumber \\
&&\mbox{}\times {\rm Tr}\,G(\br,\br,\om)\nonumber \\
&&\mbox{}+\frac{\hbar}{8\pi^2c^2}\int_0^{\infty}d\om\om^2\eps_I(\om){\rm Tr}\,G_{R}(\br,\br,\om)\nonumber \\
&&\mbox{}=\frac{\hbar}{8\pi^2c^2}{\rm Im}\int_0^{\infty}d\om\om^2\Big\{\frac{\pa}{\pa\om}[\om\eps]+\eps\Big\}
{\rm Tr}\,G(\br,\br) \nonumber \\
&&+\, u_N(\br)\nonumber \\
&=&u(\br)+u_N(\br),
\label{u1}
\eea
where $u(\br)$ is the energy density (\ref{enn}) obtained in Section II.

Two ``additional" contributions to the energy density have appeared naturally in the calculation of $U(\br)$ based on Poynting's theorem. The first one is the energy $-R_{\rm abs}t$ [Eq. (\ref{rabs})] associated with the absorptive loss of field energy, which is exactly cancelled by the energy $+R_{\rm abs}t$ picked up by the reservoir oscillators, as already indicated in Eqs. (\ref{U1})-(\ref{cancelation}) in Section \ref{sec:dipoles}B. 

The second additional contribution, given by $u_N(\br)$, is exactly the same as in Eq. (\ref{enoise}):
\bea
u_N(\br)&=&\frac{\hbar}{8\pi^2c^2}\int_0^{\infty}d\om\om^2\eps_I(\br,\om){\rm Tr}\,G_{R}(\br,\br,\om) \nonumber \\
&=&\frac{1}{2}\frac{1}{8\pi} \la \bE\cdot{\hat{\rm \bf K}} + {\hat{\rm \bf K}\cdot\bE} \ra .
\eea
As in the discussion that led to Eq. (\ref{enoise}), this shows that 
$u_N(\br)$ is  the polarization energy acquired by the ``noise" dipole moments of the atoms
as they interact with the electromagnetic field. 


\subsection{Homogeneous Medium}

When spatial dispersion is negligible, equation (\ref{u1}) is a general expression for the zero-point energy density of a linear dielectric medium, including the effects of the fluctuating electric and magnetic fields, the material medium, and the reservoir. We now show that this expression reduces to the expected energy density in the limit of a 
homogeneous medium.

The (retarded) Green dyadic for a homogeneous dielectric medium is given by \cite{bigjulie}
\be
\overline{G}_{ij}(\br,\br',\om)=\Big(1+\frac{1}{k^2_m}\nabla\nabla\Big)\frac{e^{ik_m|\br-\br'|}}{|\br-\br'|},
\ee
where $k^2_m=\eps(\om)\om^2/c^2$. Thus,  
\bea
\overline{G}_{ij}(\br,\br',\om)&=&-\frac{4\pi}{3k_m^2}\delta_{ij}\delta^3(\br-\br') \nonumber \\
&&+\Big[\delta_{ij}-R_jR_j/R^2-(\delta_{ij}-3R_iR_j/R^2)\nonumber \\
&&\mbox{}\times(\frac{1}{k_m^2R^2}-\frac{i}{k_mR})\Big]\frac{e^{ik_mR}}{R},
\label{green100}
\eea
where $R=|{\bf r}-{\bf r}'|$. For $R \rightarrow 0$,
\bea
&& \overline{G}_{ij}(\br,\br',\om)=-\frac{4\pi}{3k_m^2}\delta_{ij}\delta^3(\br-\br') \nonumber \\
&& + \frac{2i}{3} k_m \delta_{ij} - \frac{1}{k_m^2 R^3} (\delta_{ij} - 3 R_i R_j/R^2).
\label{greenlimitzero}
\eea
Ignoring for the moment the terms that diverge for $R \rightarrow 0$ (see below), we put
\bea
\overline{G}_{Iij}(\br,\br,\om) &=& [ 2 n_R(\omega) \omega/3c] \delta_{ij}, \nonumber \\
\overline{G}_{Rij}(\br,\br,\om) &=& - [ 2 n_I(\omega) \omega/3c] \delta_{ij}
\label{GhomRI}
\eea
in (\ref{u1}) and obtain
\bea
\overline{u}(\br)&=&\frac{\hbar}{4\pi^2c^3}\int_0^{\infty}d\om\om^3\Big\{[2\eps_R+\om\eps'_R]n_R \nonumber \\
&&\mbox{}-n_I\frac{\pa}{\pa\om}(\om\eps_I)-2n_I\eps_I\Big\}.
\eea 
Evaluating the second term in curly brackets by partial integration we obtain 
\be
\overline{u}(\br)=\frac{\hbar}{2\pi^2c^3}\int_0^{\infty}d\om\om^3n_R^2(\om)\frac{\pa}{\pa\om}[\om n_R(\om)].
\label{zpe1}
\ee
This is exactly the energy density derived in I for the special case of a homogeneous dielectric medium.

A few  other points regarding the model of a homogeneous medium may be worth noting. First, the Green 
dyadic  (\ref{green100}) for a homogeneous dielectric leads via equation (\ref{green101}) to the (Huttner-Barnett) expression (80) of I for the quantized (transverse) electric (and magnetic) field in a homogeneous dielectric, as is easily verified;
these expressions reduce to the standard ones in the limit of free space, as is shown by considering their limiting forms as $\eps_R\rightarrow 1,\eps_I\rightarrow 0$. Second,  equation (\ref{GhomRI}) and 
the zero-temperature expectation value [cf. Eq. (82) of I]
\be
\la \hat{\bf E}(\br,\om)\cdot\hat{\bf E}^{\dagger}(\br,\om')\ra=\frac{2\hbar\om^3}{\pi c^3}n_R(\om)\delta(\om-\om')
\ee
imply
\be
\la \hat{\bf E}(\br,\om)\cdot\hat{\bf E}^{\dagger}(\br,\om')\ra=\frac{\hbar}{\pi}\frac{\om^2}{c^2}
{\rm Tr}\,\overline{G}_{I}(\br,\br,\om)\delta(\om-\om'),
\label{greenident2}
\ee
consistent with (\ref{green103}).

The delta function term in (\ref{green100}) and (\ref{greenlimitzero})  has its origin in the singular divergence of $\hat{R}/R^2$ as $R\rightarrow 0$,
\be
\frac{1}{k_m^2}\nabla_i\nabla_j\frac{1}{|\br-\br'|}= -\frac{4\pi}{3k_m^2}\delta_{ij}\delta^3(\br-\br') .
\label{green200}
\ee
Physically, this is a direct consequence of the continuum and (spatial) local approximation, and can be remedied by allowing for spatial dispersion or/and the granular structure of matter \cite{Narayanaswamy10}.   As remarked earlier, such a singularity is to be subtracted from $\overline{G}_{ij}$. 
Its contribution to equation (\ref{intro1}) is a force density whose integral over all the space vanishes for bounded dielectric in a vacuum, as expected. 
Also, the $1/R^3$ divergence in (\ref{greenlimitzero}) can be remedied by replacing the point dipole to which it corresponds with a dielectric
sphere of radius ${\cal R}$ and then allowing ${\cal R} \rightarrow 0$; the result is \cite{steveb}
\begin{equation}
\overline{G}_{ij}(\br,\br',\om)= - \frac{4 \pi}{3 k_m^2} \delta_{ij} \delta^3(\br-\br')
\end{equation}
for $|\br-\br'|$ infinitesimally small. Again, as noted earlier, this term is to be subtracted from $\overline{G}_{ij}$, and gives a vanishing contribution
to the force density (\ref{intro1}).

Finally we note that throughout this paper we are working, albeit formally, 
with the full (transverse plus longitudinal) electric field and Green dyadic,
as is required when the permittivity can vary with $\br$.
If we consider, for example, the spontaneous emission rate $A$ of a guest atom in a homogeneous dielectric, which is proportional to the imaginary part of $\overline{G}$, we find that, in addition to the familiar contribution equal to the free-space emission rate times the real part of the refractive index at the transition frequency $\om_0$ (and possibly also including local field corrections), $A$ has a contribution proportional to ${\rm Im}(1/k_m^2)=(c^2/\om_0^2)\eps_I(\om_0)/|\eps(\om_0)|^2$, where $\om_0$ is the transition 
frequency [cf. Eq. (\ref{green200})]. Following the approach of Barnett {\sl et al.} \cite{steveb},
this part of $A$, which is attributable to the longitudinal part of the electric field, is found to depend on an unspecified parameter $R$ related to the effective distance between the guest atom and the host medium. Without 
modifying the continuum model to allow for such a parameter, we would obtain no such contribution to $A$. The same
is true in I, where we considered only the transverse component of the field, in line with the assumption of a homogeneous, continuous medium made in that paper, where there was no allowance for the finite distances
between atoms and the near-field interactions between atoms associated with longitudinal fields.


\subsection{Parallel Plates}

Equation (\ref{intro1}) as such is not in general very useful for the calculation of Casimir forces
between arbitrarily shaped media \cite{dalvitbook}. One exception is the force between two dielectric media occupying the half-spaces $z\le 0$ and $z\ge d$ and separated by a third dielectric \cite{igor, julie}. In this case 
\begin{eqnarray}
\eps(\br,\om)\hspace{-9pt}&&= \eps_1(\om)\theta(-z)+\eps_2(\om)\theta(z-d) \nonumber \\
&&+\, \eps_3(\om)[\theta(z) - \theta(z-d)] 
\end{eqnarray}
where $\theta(z)$ is the unit step function. Then the integral over space of the force density is simplified by the fact that $\nabla\eps(\br,\om)$ involves simple delta functions in $z$, yielding
\bea
{\bf F}(d) \hspace{-9pt}&&= \int d^3 r \, {\bf f}(\br) = -\hat{\bf z} \frac{ \hbar A}{8\pi^2 c^2}{\rm Im}\int_{0}^{\infty}d\om\om^2\coth\Big(\frac{\hbar\om}{2k_BT}\Big) \nonumber \\
&& \; \times \Big\{ \eps_3(\om) G_{ii}(0^{+},0^{+},\om)  - \eps_1(\om) G_{ii}(0^{-},0^{-},\om)\nonumber \\
&&+ \; \eps_2(\om) G_{ii}(d^{+},d^{+},\om) - \eps_3(\om) G_{ii}(d^{-},d^{-},\om) \Big\} ,
\label{PP2}
\eea
where $G_{ii}(0^{\pm},0^{\pm},\om)$ and $G_{ii}(d^{\pm},d^{\pm},\om)$ stand for $G_{ii}(z\rightarrow 0^{\pm}, z'\rightarrow 0^{\pm},\om)$ and $G_{ii}(z\rightarrow d^{\pm}, z'\rightarrow d^{\pm},\om)$, respectively. The Green dyadic for this geometry can be evaluated without difficulty \cite{igor, Greenmultilayer}, and its trace at coincidence can be written as
\bea
G_{ii}(z,z,\om) \hspace{-8pt}&&= \frac{i}{2\pi} \frac{c^2}{\om^2 \eps_1} \int \frac{d^2 {\bf k}_{\|}}{\kappa_1}
\Bigg\{ \left[ -\kappa_1^2 + k_{\|}^2 \right] R_1^{\rm tm} \nonumber \\
&&+ \frac{\om^2}{c^2} \eps_1  R_1^{\rm te} \Bigg\} e^{-2i \kappa_1 z} \;\;\;\; (z \leq 0) ,
\label{Gpp1}
\eea
\bea
&&G_{ii}(z,z,\om) = \frac{i}{2\pi} \frac{c^2}{\om^2 \eps_3} \int \frac{d^2 {\bf k}_{\|}}{\kappa_3} \; \times \nonumber \\
&& \Bigg\{ \frac{\kappa_3^2}{D_{\rm tm}} \left[ 2 \,r_1^{\rm tm} r_2^{\rm tm} e^{2i \kappa_3 d} - r_1^{\rm tm}e^{2i \kappa_3 z} -  r_2^{\rm tm} e^{2i \kappa_3 (d-z)}\right] \nonumber \\
&&+ \frac{\om^2}{c^2}\frac{\eps_j}{D_{\rm te}}\left[ 2\, r_1^{\rm te} r_2^{\rm te} e^{2i \kappa_3 d} + r_1^{\rm te}e^{2i \kappa_3 z} +  r_2^{\rm te} e^{2i \kappa_3 (d-z)}\right]  \nonumber \\
&&+ \frac{k_{\|}^2}{D_{\rm tm}}\left[ 2\, r_1^{\rm tm} r_2^{\rm tm} e^{2i \kappa_3 d} + r_1^{\rm tm}e^{2i \kappa_3 z} +  r_2^{\rm tm} e^{2i \kappa_3 (d-z)}\right]  \Bigg\} \nonumber \\
&& \hspace{80pt} (0 < z < d) ,
\label{Gpp3}
\eea
\bea
G_{ii}(z,z,\om) \hspace{-8pt}&&= \frac{i}{2\pi} \frac{c^2}{\om^2 \eps_2} \int \frac{d^2 {\bf k}_{\|}}{\kappa_2}
\Bigg\{ \left[-\kappa_1^2 + k_{\|}^2 \right] R_2^{\rm tm} \nonumber \\
&&+ \frac{\om^2}{c^2} \eps_2  R_2^{\rm te} \Bigg\} e^{2i \kappa_2 z} \;\;\;\; (z \geq d) ,
\label{Gpp2}
\eea
where $\kappa_j = \sqrt{\eps_j \omega^2/c^2 - k_{\|}^2}$ (the $\om$-dependence of $\eps_j$ is implicit throughout),
the Fresnel transverse electric (te) and transverse magnetic (tm) reflection coefficients are given by
\bea
r_{j}^{\rm te} = \frac{\kappa_3 - \kappa_j}{\kappa_3 - \kappa_j} \;\;\;\; , \;\;\;\;  r_{j}^{\rm tm} = \frac{\eps_j \kappa_3 - \kappa_j}{\eps_j \kappa_3 - \kappa_j} ,
\eea
and we defined
\bea
R_{1}^{\rm p} = \frac{-r_{1}^{\rm p} + r_{2}^{\rm p}e^{ 2i \kappa_3 d}}{1 - r_{1}^{\rm p} r_{2}^{\rm p}e^{2i \kappa_3 d} }  , 
\eea
\be
R_{2}^{\rm p} =  \frac{-r_{2}^{\rm p} + r_{1}^{\rm p}e^{ 2i \kappa_3 d}}{1 - r_{2}^{\rm p} r_{1}^{\rm p}e^{2i \kappa_3 d} } ,
\ee
\bea
D_{\rm p} = 1 - r_{1}^{\rm p} r_{2}^{\rm p}e^{2i \kappa_{3} d} \;\;\;\;\;\; ({\rm p} = {\rm te, tm}) .
\eea
Substituting (\ref{Gpp1})-(\ref{Gpp2}) into (\ref{PP2}) we get, after straightforward (if a bit long) manipulations
\bea
{\bf F}(d)\hspace{-8pt}&& = \int d^3 r \, {\bf f}(\br) = - \hat{\bf z} \frac{\hbar A}{4\pi^3} \, {\rm Im}  \int_{0}^{\infty} d\om \coth\left(\frac{\hbar\om}{2k_BT}\right) \nonumber \\
&&\times \int d^2 {\bf k}_{\|} \, \kappa_3 \sum_{{\rm p}={\rm te},{\rm tm}} \frac{r_{1}^{\rm p} r_{2}^{\rm p} \, e^{2i \kappa_3 d}}{D_{\rm p}} \, ,
\label{PPfinal}
\eea
that is just the force between two parallel half-spaces separated by a dielectric \cite{igor}.


\section{Casimir Force Density from the Maxwell Stress Tensor} \label{sec:stress}
The force on the material resulting from the Lorentz forces acting on the particles constituting the medium can
be obtained from the Maxwell stress tensor \cite{felder}. We now show that this force is equal to the Casimir force density
(\ref{intro1}).

Recall first the classical theory for the force density in a dielectric medium in which there are no free
charges or currents. From the macroscopic Maxwell equations one derives a force density $f_i=\pa_jT_{ij}$,
where the stress tensor is
\be
T_{ij}=\frac{1}{4\pi}[E_iD_j+H_iH_j-\frac{1}{2}({\bf E}\cdot{\bf D}+{\bf H}\cdot{\bf H})\delta_{ij}].
\label{minkeq}
\ee
It follows that
\be
f_i=\frac{1}{8\pi}[(\pa_iE_j)D_j-E_j(\pa_iD_j)].
\label{force100}
\ee
In identifying ${\bf f}(\br)$ as the force per unit volume on the medium at the point $\br$ we are assuming
conditions such that $(\pa/\pa t)({\bf D}\times{\bf H})/4\pi c$ can be taken to be zero, since we are dealing with a
stationary equilibrium situation (see also  \cite{mink}). We continue to restrict ourselves here, as in I, to isotropic media,
in which case the Minkowski form of the stress tensor defined by (\ref{minkeq}) is symmetric, consistent with
angular momentum conservation.

In quantum theory we replace $E_j$ and $D_j$ in (\ref{force100}) by operators, symmetrize, and take expectation
values:
\be
f_i=\frac{1}{8\pi}{\rm Re}\Big[\la(\pa_i\hat{E}_j)\hat{D_j}\ra-\la\hat{E}_j(\pa_i\hat{D}_j)\ra\Big],
\ee
or
\bea
f_i(\br)&=&\frac{1}{8\pi}{\rm Re}\int_{-\infty}^{\infty}d\om\int_{-\infty}^{\infty}d\om'\Big
\{\la[\pa_i\hat{E}_j(\br,\om)]\hat{D}_j(\br,\om')\ra
\nonumber \\
&&\mbox{}-\la\hat{E}_j(\br,\om)[\pa_i\hat{D}_j(\br,\om')]\ra\Big\}.
\eea
Since the electric displacement vector $\bD(\br,\om)=\eps(\br,\om)\bE(\br,\om)+\bK(\br,\om)$,  
\bea
\la\hat{E}_j(\br,\om)\hat{D}_j(\br',\om')\ra&=&\eps(\br',\om')\la\hat{E}_j(\br,\om)\hat{E}_j(\br',\om')\ra\nonumber \\
&&\mbox{}+\la\hat{E}_j(\br,\om)\hat{K}_j(\br',\om')\ra.
\eea
In the second term on the right we use (\ref{green101}), while in the first term we use (\ref{green103}).
Then
\bea
{\rm Re}\la\hat{E}_j(\br,\om)\hat{D}_j(\br',\om')\ra&=&\frac{\hbar\om^2}{\pi c^2}{\rm Im}
[\eps(\br',\om)G_{jj}(\br,\br',\om)]\nonumber \\
&&\mbox{}\times\delta(\om-\om'),
\eea
which leads us again to the force density (\ref{intro1}):
\bea
f_i(\br)&=&\frac{1}{8\pi}{\rm Im}\int_{0}^{\infty}d\om\int_{0}^{\infty}d\om'\nonumber \\
&&\mbox{}\times\lim_{\br'\rightarrow\br}\Big
\{\la[\pa_i\hat{E}_j(\br,\om)]\hat{D}_j(\br',\om')\ra
\nonumber \\
&&\mbox{}-\la\hat{E}_j(\br',\om)[\pa_i\hat{D}_j(\br,\om')]\ra\Big\}\nonumber\\
&=&-\frac{\hbar}{8\pi^2c^2}{\rm Im}\int_{0}^{\infty}d\om\om^2\pa_i\eps(\br,\om)G_{jj}(\br,\br,\om).\nonumber \\
\eea

\section{Concluding Remarks}\label{sec:conclusions}
In Section \ref{sec:dipoles} we considered a collection of stationary electrically polarizable particles in free space 
and at zero temperature and obtained formal expressions for the many-body van der Waals interaction energies among these particles. The results are equivalent to those obtained by various other methods, but unlike most of earlier work we accounted explicitly for dissipation by coupling each particle to a reservoir of harmonic oscillators rather than by invoking analytic properties of the permittivity. Within the continuum approximation we obtained expressions for the electromagnetic energy density and a Casimir force density; as shown in the Appendix, these expressions are easily
generalized to the case of thermal equilibrium. The force density obtained is in agreement with that originally derived by Dzyaloshinskii {\em et al.} \cite{igor}. In Section \ref{sec:macro} we derived the formula (\ref{enn}) for the energy density starting from the continuum approximation and Poynting's theorem in macroscopic quantum electrodynamics, and in Section \ref{sec:stress} we derived the formula (\ref{intro1}) for the force density using the Minkowski form of the stress tensor.

Under the assumption of a continuous medium, and ignoring single-particle self-energies, the expression (\ref{enn}) is {\em exact} \cite{philbin}. Similar expressions are well known in classical electromagnetic theory \cite{landaujack} as {\em approximations} applicable for frequencies at which absorption is negligible. The reason our expression is exact is not, strictly speaking, due to our quantum treatment of the field, but rather to the fact that Fourier components of the zero-point field field at different frequencies are uncorrelated \cite{pap1} (as they are at finite equilibrium temperatures). 

Under the (unrealistic) assumption that absorption can be completely ignored, so that $\eps$ is real, we would write
\bea
u(\br)&=&\frac{1}{8\pi}\int_0^{\infty}d\om\Big\{\frac{\pa}{\pa\om}[\om\eps(\br,\om)]+\eps(\br,\om)\Big\}\nonumber \\
&&\mbox{}\times\la\bE(\br,\om)\cdot\bE^{\dag}(\br,\om)\ra\nonumber \\
&=&\frac{\hbar}{8\pi^2c^2}{\rm Im}\int_0^{\infty}d\om\om^2\Big\{\frac{\pa}{\pa\om}[\om\eps(\br,\om)]+\eps(\br,\om)\Big\}\nonumber \\
&&\mbox{}\times G_{ii}(\br,\br,\om).
\label{noabs}
\eea
Since both $\eps$ and $G$ as functions of complex frequency are analytic in the upper half of the complex frequency
plane, we can perform a Wick rotation and write
\bea
u(\br)&=&\frac{\hbar}{8\pi^2c^2}{\rm Im}\int_0^{\infty}d\xi\xi^2\Big\{\frac{\pa}{\pa\xi}[\xi\eps(\br,i\xi)]+\eps(\br,i\xi)\Big\}\nonumber \\
&&\mbox{}\times G_{ii}(\br,\br,i\xi),
\eea
where we have used the relation between $\la\bE(\br,\om)\cdot\bE^{\dag}(\br,\om)\ra$ and $G(\br,\br,\om)$ for
purely imaginary frequencies $i\xi$. In other words, consistent with the remarks in the Introduction, we
can obtain the correct energy density from the formula (\ref{noabs}), which assumes there is no absorption, by
formally replacing $\eps(\br,\om)$ and $G(\br,\br,\om)$ by their (real) values $\eps(\br,i\xi)$ and $G(\br,\br,i\xi)$.
Similarly we can formally interpret the energy density (\ref{enn}) as an integral over zero-point energies $(1/2)\hbar\nu$
by writing it as
\be
u(\br)=\int_0^{\infty}d\nu\rho(\br,\nu)\Big(\frac{1}{2}\hbar\nu\Big) ,
\ee
where 
\be
\rho(\br,\nu)=\frac{\nu}{4\pi^2c^2}\Big\{\frac{\pa}{\pa\nu}[\nu\eps(\br,i\nu)]+\eps(\br,i\nu)]\Big\}
\ee
is a local density of states \cite{densityofstates}.

\section*{Acknowledgments}
We thank G. Barton, L. S. Brown, S. Y. Buhmann, I. E. Dzyaloshinskii, J.-J. Greffet, F. Intravaia, A. Narayanaswamy, and T. G. Philbin  for discussions relating to this work. We acknowledge funding by DARPA/MTO's Casimir Effect Enhancement program under DOE/NNSA Contract No. DE-AC52-06NA25396.
This research was also partially supported by Triangle de la Physique, under the contract 2010-037T-EIEM.


\section*{Appendix. Generalization to Finite Temperature}
It is an easy matter to generalize our approach and results to finite equilibrium temperatures $T$. In the approach of
Section \ref{sec:dipoles} we replace $\la{\bf E}(\br)\cdot{\bf E}^{\dag}(\br)\ra$ in Eq. (\ref{hf1}) by
$\la{\bf E}(\br)\cdot{\bf E}^{\dag}(\br)+{\bf E}^{\dag}(\br)\cdot{\bf E}(\br)\ra$, since the quantized field ${\bf E}$ 
involves photon annihilation operators and does not give zero when acting on a thermal state. We 
similarly replace (\ref{hf2}) by \cite{greenident}
\bea
\la\bE_0(\br,\om)\cdot\bE^{\dag}_0(\br,\om')+\bE^{\dag}_0(\br,\om)\cdot\bE_0(\br,\om')\ra=\nonumber\\
\frac{\hbar}{\pi}\coth\Big(\frac{\hbar\om}{2k_BT}\Big)
{\rm Im}{\rm Tr}\,G_{0}(\br,\br,\om)\delta(\om-\om').
\eea
Then formulas such as (\ref{hf4})-(\ref{heq4}) and (\ref{enn}) are generalized to finite temperature by simply 
inserting the factor $\coth(\hbar\om/2k_BT)$ in the integrations over $\om$. Thus, for example, the energy
density (\ref{enn}) generalizes to
\bea
{u}(\br) &=&\frac{\hbar}{8\pi^2c^2}{\rm Im}{\rm Tr}\int_0^{\infty}d\om\om^2
\coth\Big(\frac{\hbar\om}{2k_BT}\Big)\nonumber\\
&&\mbox{}\times[2\eps(\br,\om)+\om\eps'(\br,\om)]G(\br,\br,\om).
\label{ennT}
\eea
The same arguments leading from the energy density to the force density apply regardless of the temperature, so that the generalization of (\ref{intro1})
is simply
\bea
{\bf f}(\br)&=&-\frac{\hbar}{8\pi^2 c^2}{\rm Im}\int_{0}^{\infty}d\om\om^2\coth\Big(\frac{\hbar\om}{2k_BT}\Big)\nabla\eps(\br,\om)\nonumber\\
&&\mbox{}\times G_{ii}(\br,\br,\om).
\label{intro1T}
\eea
We can convert in the standard fashion the integral over the real $\om$ axis to an integral over the imaginary $\om$ axis. Using the fact that $\eps(\br,\om)$ and $G(\br,\br,\om)$ are analytic in the upper half of the complex
frequency plane, and that $\coth(\hbar\om/2k_BT)$ has poles at $\om_n=(2\pi ik_BT/\hbar)n=i\xi_n$ for all integers
$n$, the result is that the integration over $\om$ is replaced by a sum over $n$:
\be
\int_0^{\infty}d\om f(\om)\rightarrow -\frac{2\pi k_BT}{\hbar} \sum_{n=0}^{\infty'}f(i\xi_n),
\ee
where $f(\om)$ denotes the integrand in (\ref{intro1T}) and the prime on the summation sign indicates that 
the $n=0$ term is multiplied by 1/2. Then
\be
{\bf f}(\br)=-\frac{k_BT}{4\pi c^2} \sum_{n=0}^{\infty'}  \om_n^2G_{ii}(\br,\br,i\om_n)\nabla\eps(\br,i\om_n),
\label{dd1T}
\ee
which is the result of Dzyaloshinskii \emph{et al.} quoted earlier [Eq. (\ref{dd1})]. The same finite-temperature results are easily shown to emerge from the approach of Section \ref{sec:macro} when the temperature-dependent terms in the noise polarization correlation functions (\ref{prop1}) and (\ref{prop2}) are retained.



\begin{thebibliography}{99} 

\bibitem{fieldquant} Among the many papers on the subject are: V.L. Ginzburg, J. Phys., USSR, \textbf{2}, 441 (1940) and \textbf{3}, 101 (1940); J.M. Jauch and K.M. Watson, Phys. Rev. \textbf{74}, 950 (1948); M. Hillery and
L.D. Mlodinow, Phys. Rev. A \textbf{30}, 1860 (1984); P.D. Drummond, Phys. Rev. A \textbf{42}, 6845 (1990);
R.J. Glauber and M. Lewenstein, Phys. Rev. A \textbf{43}, 467 (1991); P.W. Milonni, J. Mod. Opt. \textbf{42}, 1991 (1995).
For our purposes here it suffices to take the magnetic permeability to be unity at all
frequencies. It is straightforward to extend the macroscopic quantization approach to the more general case of magnetodielectric and possibly negative-index media; see, for instance, P.W. Milonni and G.J. Maclay, Opt. Commun.
{\bf 228}, 161 (2003), where this is done for frequencies at which absorption is negligible. 

\bibitem{huttnerbarnett} B. Huttner and S.M. Barnett, Phys. Rev. A \textbf{46}, 4306 (1992). 
See also, for instance, R. Matloob, R. Loudon, and S.M. Barnett, Phys. Rev. A {\bf 52}, 4823 (1995); 
H.T. Dung, L. Kn\"oll, and D.-G. Welsch, Phys. Rev. A {\bf 57}, 3931 (1998);
N.A.R. Bhat and J.E. Sipe, Phys. Rev. A {\bf 73}, 063808 (2006) and references therein.

\bibitem{term} The homogeneous media we refer to are also assumed throughout to be isotropic. 

\bibitem{fano} U. Fano, Phys. Rev. \textbf{103}, 1202 (1956).

\bibitem{Rytov} See, for example, S.M. Rytov, Sov. Phys. JETP {\bf 6}, 130 (1958).

\bibitem{lifshitz} E.M. Lifshitz, Sov. Phys. JETP {\bf 2}, 73 (1956).

\bibitem{erber} The relation between the Fano-diagonalization and Langevin-force approaches in this
context is discussed, for instance, by F.S.S. Rosa, D.A.R. Dalvit, and P.W. Milonni, arXiv:0912.0279v1 [quant-ph] and
in {\sl Doing Physics. A Festshcrift for Thomas Erber}, ed. P.W. Johnson (Illinois Institute of Technology Press, Chicago, IL, 2010).

\bibitem{pap1} F.S.S. Rosa, D.A.R. Dalvit, and P.W. Milonni, Phys. Rev. A {\bf 81} 033812 (2010).

\bibitem{pap1typos} In the line following Eq. (32) in Reference \cite{pap1}, ``the (cycle-averaged) {\em kinetic} energy" should be replaced by ``twice the (cycle-averaged) {\em kinetic} energy." Also, the first line of Eq. (107) should be
${\rm Re} \, \eps^{3/2}=(n_R^2-n_I^2)n_R-2n_I^2n_R$.

\bibitem{igor} I.E. Dzyaloshinskii and L.P. Pitaevskii, Sov. Phys. JETP {\bf 9}, 1282 (1959); I.E. Dzyaloshinskii,
E.M. Lifshitz, and L.P. Pitaevskii, Adv. Phys. {\bf 10}, 165 (1961). See also A.A. Abrikosov, L.P. Gorkov, and I.E. Dzyaloshinskii, {\sl Methods of Quantum Field Theory in Statistical Physics} (Dover, N.Y., 1975).

\bibitem{julie} J. Schwinger, L.L. DeRaad, Jr., and K.A. Milton, Ann. Phys. (N.Y.) {\bf 115}, 1 (1978). $G(\br,\br,\om)$
in Eq. (\ref{intro1}) is $4\pi c^2/\om^2$ times the Green function $\Gamma(\br,\br,\om)$ of Schwinger, DeRaad, and Milton.

\bibitem{vankamp} See, for instance, N.G. van Kampen, B.R.A. Nijboer, and K. Schram, Phys. Lett. {\bf 26}A, 307 (1968).

\bibitem{casimir} H.B.G. Casimir, Proc. K. Ned. Akad. Wet. 51, 793 (1948).

\bibitem{agarwal} G.S. Agarwal, Phys. Rev. A {\bf 11}, 243 (1975).

\bibitem{pwm} P.W. Milonni, {\sl The Quantum Vacuum. An Introduction to Quantum Electrodynamics} (Academic, San
Diego, 1994).

\bibitem{ginzburg} V.L. Ginzburg, {\sl Theoretical Physics and Astrophysics} (Pergamon, Oxford, 1979), p. 309.

\bibitem{fordkac} For a discussion of such a quantum Langevin equation see G. W. Ford and M. Kac, J. Stat. Phys. {\bf
46}, 803 (1987) and references therein. 

\bibitem{hellfeyn} See, for instance, A.L. Fetter and J.D. Walecka, {\em Quantum Theory of Many-Particle Systems} (McGraw-Hill, N.Y., 1971), pp. 69-70.

\bibitem{greenident} See, for instance, G.S. Agarwal, Phys. Rev. A {\bf 11}, 230 (1975); W. Eckhardt,
Phys. Rev. A {\bf 29}, 1991 (1984); M.S. Toma$\check{s}$, Phys. Rev. A {\bf 66}, 052103 (2002), 
and references therein.

\bibitem{pwm3} See, for instance, Reference \cite{pwm}, pp. 242-244.

\bibitem{stark} The expression with $\gamma=0$, based on linear response theory for the field fluctuations, appears in Reference \cite{agarwal}, Eq. (2.19). See also, for instance, Reference \cite{pwm}, Eq. (7.120). 

\bibitem{mahan} G.D. Mahan, J. Chem. Phys. {\bf 43}, 1569 (1965).

\bibitem{renne} M.J. Renne, Physica {\bf 53}, 193 (1971); {\bf 56}, 125 (1971).

\bibitem{dimen} For dimensional purposes it must be kept in mind that the formal expressions we employ for
notational conciseness imply different volume integrations. In Eq. (\ref{heq444}),
for instance, ${\rm Tr}[G_0]^n$ involves an $n$-fold volume integration, as in Eq. (\ref{heq34}) for $n=2$. In 
Eq. (\ref{heq6}), however, Tr involves an integration over all points in space, excluding $\br$, appearing in the argument of $G$, so that (\ref{heq444}) is an energy whereas (\ref{heq6}) is an energy density.

\bibitem{tracelog} See, for instance, K.A. Milton, {\sl The Casimir Effect. Physical Manifestations of Zero-Point
Energy} (World Scientific, Singapore, 2001).

\bibitem{milton} K.A. Milton, J. Wagner, P. Parashar, and I. Brevik, Phys. Rev. D {\bf 81}, 065007 (2010). 

\bibitem{tip} For media with piecewise constant complex permittivities the singular parts of $G(\br,\br,\om)$ and
$\overline{G}(\br,\br,\om)$ have been shown to coincide in each region of constant permittivity. See C.A. Gu\'erin,
B Gralak, and A. Tip, Phys. Rev. E {\bf 75}, 056601 (2007).

\bibitem{julians} This identity is also employed by Mahan \cite{mahan} and Renne \cite{renne} and, in a different
context, by J. Schwinger, Phys. Rev. {\bf 93}, 615 (1954), Eq. (34). 

\bibitem{BarashGinzburg} For a discussion of this point, see Yu. S. Barash and V. L. Ginzburg,
Sov. Phys. Usp. {\bf 18} 305 (1975) [Usp. Fiz. Nauk. {\bf 116} 5 (1975)].

\bibitem{rosenfeld} L. Rosenfeld, {\sl Theory of Electrons} (North-Holland, Amsterdam, 1951), p. 78.

\bibitem{bigjulie} H. Levine and J. Schwinger, Commun. Pure Appl. Math. {\bf 3}, 355 (1950).

\bibitem{Narayanaswamy10} A. Narayanaswamy and G. Chen, J. Quant. Spectrosc. Rad. Trans. {\bf 111}, 1877 (2010).

\bibitem{steveb} See, for instance, S.M. Barnett, B. Huttner, R. Loudon, and R. Matloob, J. Phys. B: At. Mol. 
Opt. Phys. {\bf 29}, 3763 (1996). For a somewhat similar analysis for the case of a dipole in a conducting medium, see C.T. Tai and R.E. Collin, IEEE Trans. Ant. Prop. {\bf 48}, 1501 (2000).  

\bibitem{dalvitbook} Computational methods for evaluating Casimir forces for arbitrary geometries are described
by several authors in {\sl Casimir Physics}, Lecture Notes in Physics Vol. {\bf 834}, eds. D. A. R. Dalvit, P. W. Milonni, D. C. Roberts, and
F. S. S. Rosa  (Springer-Verlag, Heidelberg, 2011).

\bibitem{Greenmultilayer} See, for instance, M.S. Toma$\check{s}$, Phys. Rev. A {\bf 66}, 052103 (2002).

\bibitem{mink} $(\pa/\pa t)({\bf D}\times{\bf H})/4\pi c$ is the rate of change of the momentum density of the
field in the Minkowski formulation of electromagnetic forces in dielectric media. This momentum density differs
from the more widely accepted (Abraham) form of the momentum density of the field. This distinction is not
directly relevant to our treatment here based on (\ref{force100}), which follows directly from the macroscopic Maxwell equations. Recent reviews of the Abraham-Minkowski ``controversy" include S. M. Barnett and R. Loudon, Phil. Trans. R.  Soc. A \textbf{368}, 927-939 (2010); C. Baxter and R. Loudon, J. Mod. Opt. \textbf{57}, 830 (2010); P. W. Milonni and R. W. Boyd, Adv. Opt. Phot. {\bf 2}, 519 (2010).

\bibitem{felder} The calculation in this section is similar to that of B.U. Felderhof, J. Phys. C: Solid State Phys. 
{\bf 10}, 4605 (1977), although in that work the fields appear to be treated as classical stochastic fields. We thank Prof. A. Narayanaswamy for bringing Felderhof's work to our attention.

\bibitem{philbin} T. G. Philbin [New J. Phys. {\bf 12} 123008 (2010), New J. Phys. {\bf 13}, 063026 (2011)] has recently obtained similar results by a quite different route based on a canonical quantization procedure for macroscopic electrodynamics. Dissipation
is included by invoking the analytical properties of the electric permittivity and magnetic permeability (Kramers-Kronig relations). His expressions (96) and (97) for the energy density reduce to our expressions (\ref{ennT}) and (\ref{enn}), respectively, when the medium is assumed to be purely dielectric. We thank Dr. Philbin for bringing this work to our attention.

\bibitem{landaujack} See, for instance, L.D. Landau and E.M. Lifshitz, {\em Electrodynamics of Continuous Media} (Pergamon, Oxford, 1960), Sec. 61.

\bibitem{densityofstates} For an analysis of the local density of states relating to 
dyadic electromagnetic Green functions see \cite{Narayanaswamy10}.

\end{thebibliography}
\end{document}